\documentclass[aps,prb,reprint,preprintnumbers,superscriptaddress,amsmath,amssymb,bibnotes,longbibliography]{revtex4-2}
\usepackage{graphicx}
\usepackage{dcolumn}
\usepackage{epstopdf}
\usepackage{braket}
\usepackage{bm,multirow,threeparttable}
\usepackage{flushend}
\usepackage[pdfstartview=FitH, CJKbookmarks=true, bookmarksnumbered=true, bookmarksopen=true, colorlinks, pdfborder=001, linkcolor=blue, anchorcolor=blue, citecolor=blue,urlcolor=blue]{hyperref}

\begin{document}
\title{Magnetic phases of Kondo lattice materials Ce$_5$RhGe$_2$ and Ce$_5$IrGe$_2$}
\author{Jiawen Zhang}
\affiliation  {New Cornerstone Lab, Center for Correlated Matter and School of Physics, Zhejiang University, Hangzhou 310058, China}
\author{Yanan Zhang}
\affiliation  {New Cornerstone Lab, Center for Correlated Matter and School of Physics, Zhejiang University, Hangzhou 310058, China}
\author{Mingyi Wang}
\affiliation  {New Cornerstone Lab, Center for Correlated Matter and School of Physics, Zhejiang University, Hangzhou 310058, China}
\author{Ye Chen}
\affiliation  {New Cornerstone Lab, Center for Correlated Matter and School of Physics, Zhejiang University, Hangzhou 310058, China}

\author{Yu Liu}
\affiliation  {New Cornerstone Lab, Center for Correlated Matter and School of Physics, Zhejiang University, Hangzhou 310058, China}
\author{Yongjun Zhang}
\affiliation  {Institute for Advanced Materials, Hubei Normal University, Huangshi 435002, China}
\author{Michael Smidman}
\email[Corresponding author: ]{msmidman@zju.edu.cn}
\affiliation  {New Cornerstone Lab, Center for Correlated Matter and School of Physics, Zhejiang University, Hangzhou 310058, China}
\author{Huiqiu Yuan}
\email[Corresponding author: ]{hqyuan@zju.edu.cn}
\affiliation  {New Cornerstone Lab, Center for Correlated Matter and School of Physics, Zhejiang University, Hangzhou 310058, China}
\affiliation  {Institute of Fundamental and Transdisciplinary Research, Zhejiang University, Hangzhou 310058, China}
\affiliation  {Institute for Advanced Study in Physics, Zhejiang University, Hangzhou 310058, China}
\affiliation  {State Key Laboratory of Silicon and Advanced Semiconductor Materials, Zhejiang University, Hangzhou 310058, China}
\affiliation  {Collaborative Innovation Center of Advanced Microstructures, Nanjing 210093, China}

\date{\today}

\begin{abstract}

Single crystals of Ce$_5$RhGe$_2$ and Ce$_5$IrGe$_2$ have been systematically investigated by electrical resistivity, specific heat, and magnetization measurements. Together with Ce$_5$CoGe$_2$, all three compounds crystallize in the orthorhombic \emph{Pnma} structure, with the lattice parameters increasing monotonically from Co to Rh to Ir, consistent with the effect of negative chemical pressure. Magnetization measurements along the three principal crystallographic axes identify the \emph{a} axis as the easy magnetization direction throughout the series. Ce$_5$RhGe$_2$ exhibits ferromagnetic ordering with a Curie temperature of approximately 11.5 K and shows magnetic behavior closely resembling that of Ce$_5$CoGe$_2$. In contrast, Ce$_5$IrGe$_2$ undergoes two successive magnetic transitions at $T_{\rm M1}=12.7$ K and $T_{\rm M2}=11.8$ K, and there are multiple metamagnetic transitions under magnetic fields, giving rise to magnetization plateaus at fractions of the saturation magnetization $M_{\rm s}$ of approximately $M_{\rm s}/5$ and $M_{\rm s}/3$. 
The low-field metamagnetic transition along the easy axis shifts to lower field with decreasing temperature, and eventually a pronounced hysteresis loop is observed about zero-field, establishing that Ce$_5$IrGe$_2$ exhibits a ferrimagnetic ground state at the lowest measured temperatures.

\begin{description}
\item[PACS number(s)]
\end{description}
\end{abstract}

\maketitle

\section{Introduction}

Heavy-fermion compounds, in which the competition between Kondo screening of localized 4$f$ moments and Ruderman--Kittel--Kasuya--Yosida (RKKY) intersite exchange gives rise to diverse magnetic ground states and quantum critical phenomena, have long served as important platforms for studying strongly correlated electron physics \cite{weng2016multiple,Si2010,GegenwartP2008}. 
Ferromagnetic (FM) heavy-fermion compounds have attracted particular interest because they offer an opportunity to explore different forms of quantum criticality associated with the suppression of magnetic order \cite{Brando2016,SteppkeA2013,Aoki2014superconductivity}. However, continuous FM quantum critical points (QCPs) are exceptionally rare compared with their antiferromagnetic (AFM) counterparts, because the suppression of ferromagnetism is often interrupted by competing instabilities, including fluctuation-driven first-order transitions, modulated magnetic states such as spin-density-wave or spiral order, or Kondo cluster glass \cite{Brando2016,Uhlarz2004,FriedemannS2018,Kote2013,Wester2009}. Hydrostatic pressure provides a clean means of tuning the $f$–conduction-electron hybridization in Ce-based Kondo lattices without introducing chemical disorder \cite{weng2016multiple}.
The recent discovery of pressure-induced quantum criticality in the easy-plane ferromagnetic Kondo lattice CeRh$_6$Ge$_4$ provided a rare realization of a FM QCP in a clean stoichiometric material, posing the question as to its underlying mechanism \cite{ShenB2020,Zhan2025}. 

Recently, the Kondo-lattice compound Ce$_5$CoGe$_2$ was revealed to exhibit a distinct scenario for superconductivity in proximity to a quantum phase transition \cite{Zhangyn2026}.
While Ce$_5$CoGe$_2$ undergoes a FM transition at ambient pressure
\cite{dajun2024,pikul2014magnetic}, hydrostatic pressure first induces
an AFM phase, whose ordering temperature is continuously suppressed to
zero at an AFM QCP near $P_{\rm c}\approx3.2$ GPa. Superconductivity
emerges only above approximately 6.2 GPa, well beyond the AFM QCP,
and $T_{\rm SC}$ increases to about 2 K at 15 GPa
\cite{Zhangyn2026}.
Neutron diffraction measurements revealed a non-collinear FM structure \cite{Wu2026}, where the magnetic moments on each of the four inequivalent Ce sites are canted while retaining a net moment along the $a$ axis. This magnetic structure likely arises from the competition between the RKKY interaction, the Dzyaloshinskii--Moriya interaction, and crystalline electric field (CEF) effects.

Prior to the successful growth of single crystals of Ce$_5$CoGe$_2$, investigations of the Ce$_5M$Ge$_2$ ($M$ = transition metal) family were largely limited to polycrystalline samples \cite{sologub2000formation}. Depending on the transition metal, these compounds crystallize in either an orthorhombic Y$_5$HfS$_2$-type structure (space group $Pnma$; $M$ = Co, Rh, Ir, Ru, and Pd) or a hexagonal Mn$_5$Si$_3$-type structure (space group $P6_3/mcm$; $M$ = Ni and Ag). Among the orthorhombic members, Ce$_5$RhGe$_2$ exhibits FM ordering with $T_{\rm C}\approx11.5$ K \cite{skornia2019magnetic}, whereas Ce$_5$PdGe$_2$ and Ce$_5$RuGe$_2$ display ferrimagnetic (FIM) ground states accompanied by field-induced metamagnetic transitions \cite{skornia2017electronic,skornia2018electronic}. For the Ir analogue Ce$_5$IrGe$_2$, measurements of polycrystalline samples suggested an AFM transition near 12.5 K \cite{sologub2000formation}. Establishing the ambient-pressure magnetic ground states of these materials and their field-evolution is therefore essential for elucidating how magnetism evolves under pressure and how it is connected to the emergence of quantum criticality and superconductivity.

In this work, we report the successful growth of single crystals of Ce$_5$RhGe$_2$ and Ce$_5$IrGe$_2$, and investigate their physical properties as a function of temperature and magnetic field. Ce$_5$RhGe$_2$ exhibits FM behavior closely resembling that of Ce$_5$CoGe$_2$. Ce$_5$IrGe$_2$ displays a qualitatively different magnetic evolution characterized by two successive magnetic transitions, multiple metamagnetic transitions, and fractional magnetization plateaus at approximately $M_{\rm s}/5$ and $M_{\rm s}/3$.
We establish the magnetic field–temperature ($H$-$T$) phase diagram of Ce$_5$IrGe$_2$, revealing that the low-field metamagnetic transition shifts toward lower fields upon cooling, while a pronounced zero-field hysteresis develops at the lowest temperatures, establishing a FIM-like ground state.

\section{Experimental methods}

Single crystals of Ce$_5$$M$Ge$_2$ ($M$ = Co, Rh, Ir) were grown by the self-flux method, exploiting the relatively low melting points of the Ce-$M$ eutectic alloys \cite{dajun2024}. 
To ensure homogeneous mixing of the constituent elements, the starting materials were first arc-melted under a high-purity argon atmosphere.
The resulting ingots were then transferred into tantalum crucibles, sealed inside evacuated quartz ampules, and placed in a muffle furnace. The assemblies were heated to 1150 $^{\circ}$C, held at this temperature for 24 h, and subsequently cooled at a rate of 2 $^{\circ}$C/h, to temperatures of 550, 750, and 700 $^{\circ}$C, for $M$ = Co, Rh, and Ir, respectively. After centrifugation to remove the excess flux, shiny, rectangular-shaped single crystals were successfully obtained.

The actual chemical compositions were checked by energy-dispersive x-ray spectroscopy (EDS) with a Hitachi SU-8010 field emission scanning electron microscope. The measured atomic ratios of Ce:$M$:Ge were 5:1:2 with no significant elemental deficiency. The crystal structure was characterized by single-crystal x-ray diffraction (XRD) using a Bruker D8 Venture diffractometer with Mo K$_\alpha$ radiation. Resistivity measurements were performed in a Physical Property Measurement System (PPMS, Quantum Design) using a standard four-probe technique. The specific heat was measured by the relaxation method on the same PPMS platform. The magnetization and magnetic susceptibility measurements were performed using a Quantum Design Magnetic Property Measurement System (MPMS-5T) from 1.8 K to 300 K.

\section{Results}

\begin{figure}[htbp]
	\begin{center}
		\includegraphics[width=\columnwidth]{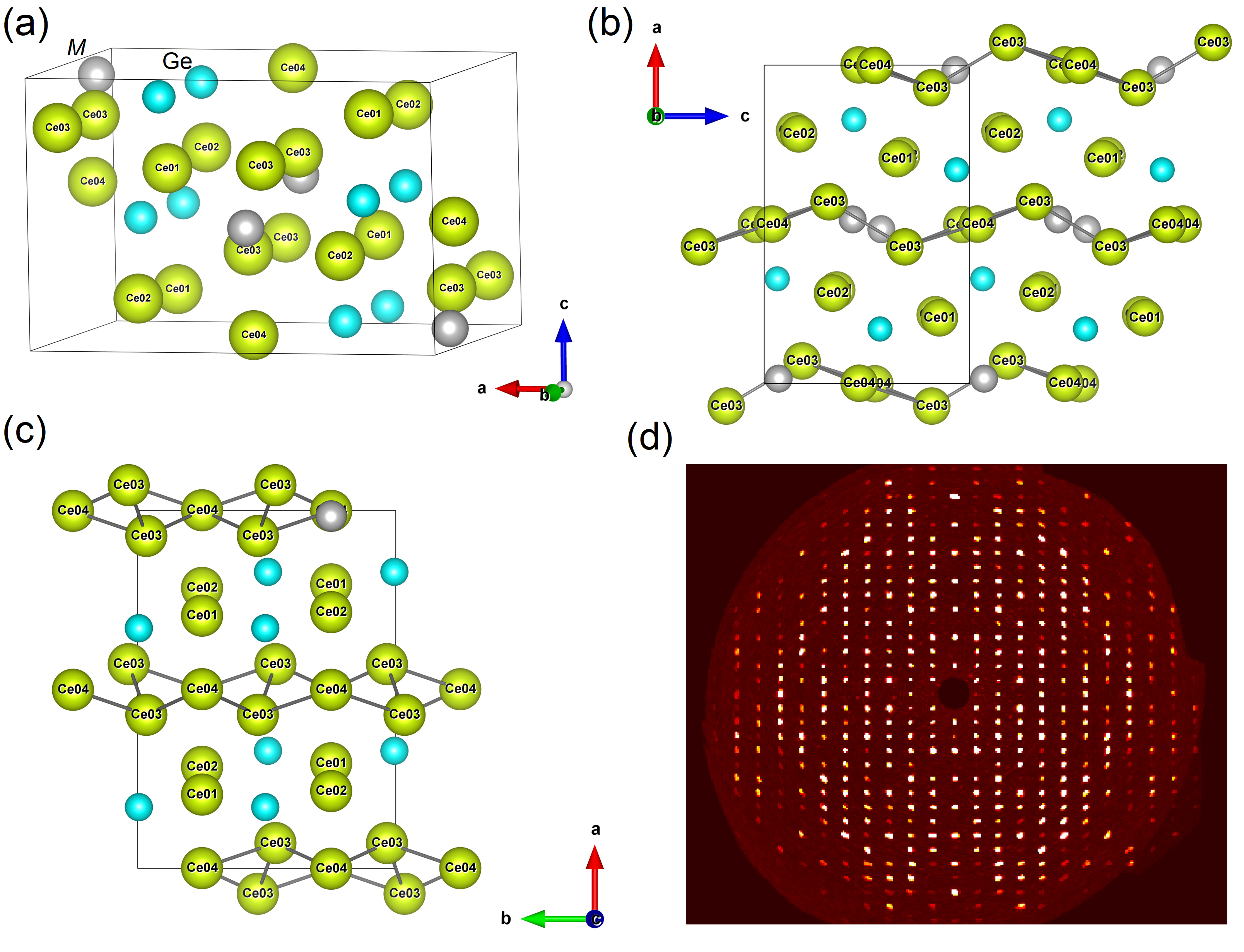}
	\end{center}
	\caption{(Color online) (a)-(c) The structure of Ce$_5$$M$Ge$_2$ ($M$ = Co, Rh, Ir), showing the Ce atomic arrangement with Ce--Ce distances shorter than 3.55 \AA{} highlighted. Green, gray, and blue spheres represent Ce, $M$, and Ge atoms, respectively. (d) Single crystal XRD pattern of Ce$_5$IrGe$_2$ in the ($h$ 0 $l$) plane.}
	\label{figure1}
\end{figure}

Refinement of single-crystal XRD data confirms that all three compounds crystallize in the orthorhombic space group $Pnma$ (No. 62), with the refined lattice parameters summarized in Tables \ref{table1} and \ref{table2}. Across the series from Co to Rh and Ir, both the lattice constants and unit-cell volumes exhibit monotonic increases, consistent with the substitution of larger transition-metal ions. 
The crystal structure of Ce$_5$$M$Ge$_2$ is illustrated in Figs. \ref{figure1}(a)–(c) and contains four crystallographically inequivalent Ce sites.
The nearest-neighbor Ce(3)–Ce(4) and Ce(3)–Ce(3) bonding networks are highlighted to illustrate the underlying Ce sublattice connectivity. Additionally, a representative single-crystal XRD pattern of Ce$_5$IrGe$_2$ is presented in Fig. \ref{figure1}(d), confirming the single-crystalline nature of the samples.

\begin{table}[!ht]
	\renewcommand\arraystretch{1.4}
	\caption{Crystallographic parameters of Ce$_5$$M$Ge$_2$ ($M$ = Co, Rh, Ir).}
	\begin{tabular}{lccc}
		\hline\hline
		 & ~~~Ce$_5$CoGe$_2$~~~ & ~~~Ce$_5$RhGe$_2$~~~ & ~~~Ce$_5$IrGe$_2$~~~
		\\ \hline
		$a$ (\AA) & 12.2684(4) & 12.3094(4) & 12.3547(3)
		\\
		$b$ (\AA) & 8.8466(3) & 8.9277(2) & 8.9294(2)
		\\
		$c$ (\AA) & 7.9155(3) & 7.9773(2) & 7.9893(2)
		\\
		$V$ (\AA$^3$) & 859.10(5) & 876.66(4) & 881.38(4)
		\\ \hline
	\end{tabular}
	\label{table1}
\end{table}

\begin{table*}[!ht]
	\renewcommand\arraystretch{1.4}
	\centering
	\caption{Structural parameters and equivalent isotropic displacement parameters  of Ce$_5$IrGe$_2$.}
	\begin{tabular}{lccccc}
		\hline\hline
		Atoms~~~ & ~~~~~~~~~~~~$x$~~~~~~~~~~~~ & ~~~~~~~~~~~~$y$~~~~~~~~~~~~ & ~~~~~~~~~~~~$z$~~~~~~~~~~~~ & ~~~~~~~~~$U_{\rm eq}$ (\AA$^2$)~~~~~~~~ & ~~~~~~~Wyck.~~~~~~~~
		\\ \hline
		Ce(1) & 0.29371(4) & 0.25 & 0.35276(7) & 0.00725(13) & 4c 
		\\
		Ce(2) & 0.71610(4) & 0.25 & 0.66795(6) & 0.00743(13) & 4c 
		\\
		Ce(3) & 0.42914(3) & 0.53405(4) & 0.68449(5) & 0.00966(11) & 8d
		\\
		Ce(4) & 0.49918(4) & 0.25 & 0.96183(7) & 0.00845(13) & 4c
		\\
		Ir & 0.48471(3) & 0.25 & 0.57001(5) & 0.01098(12) & 4c
		\\
		Ge & 0.67123(6) & 0.50519(8) & 0.93659(9) & 0.00798(16) & 8d
		\\ \hline
	\end{tabular}
	\label{table2}
\end{table*}

\begin{figure}[htbp]
	\begin{center}
		\includegraphics[width=\columnwidth]{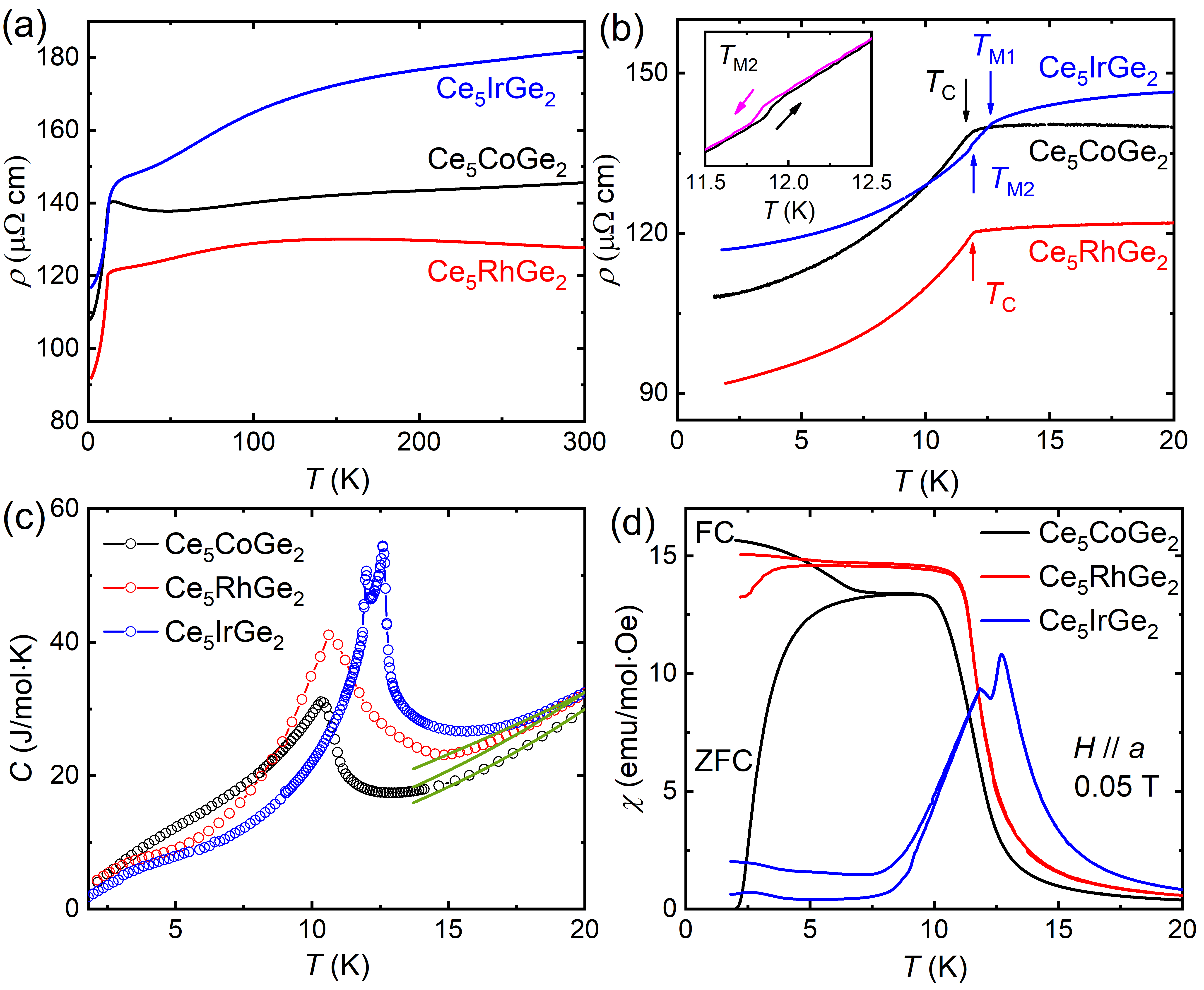}
	\end{center}
	\caption{(Color online) (a) Temperature dependence of the electrical resistivity $\rho(T)$ of Ce$_5$$M$Ge$_2$ ($M$ = Co, Rh, Ir). (b) $\rho(T)$ at low temperatures. The black and red arrows indicate the ferromagnetic transition temperatures $T_{\rm C}$ of Ce$_5$CoGe$_2$ and Ce$_5$RhGe$_2$, respectively, while the two blue arrows indicate the magnetic transition temperatures $T_{\rm M1}$ and $T_{\rm M2}$ of Ce$_5$IrGe$_2$. The inset shows the thermal hysteresis across $T_{\rm M2}$ upon heating and cooling. (c) Temperature dependence of the specific heat of Ce$_5$$M$Ge$_2$. The green lines are fits to $C(T) = \gamma_0 T + \beta T^{3}$. (d) Temperature dependence of the magnetic susceptibility $\chi(T)$ measured at 0.05 T with $H \parallel a$. Both zero-field-cooled (ZFC) and field-cooled (FC) curves are shown.}
	\label{figure2}
\end{figure}

Figure \ref{figure2}(a) displays the temperature dependence of the electrical resistivity, $\rho(T)$, of Ce$_5$$M$Ge$_2$ ($M$ = Co, Rh, Ir). At high temperatures, all three compounds exhibit a broad hump around 150 K, which is a typical feature arising from the interplay between CEF and the Kondo scattering of $4f$ electrons \cite{dajun2024}. At low temperatures [Fig. \ref{figure2}(b)], the $\rho(T)$ curves for Ce$_5$CoGe$_2$ and Ce$_5$RhGe$_2$ drop abruptly at 11 K and 11.5 K, respectively, signaling the onset of long-range magnetic order. 
For Ce$_5$IrGe$_2$, two successive anomalies are observed at $T_{\rm M1} \approx 12.7$ K and $T_{\rm M2} \approx 11.8$ K. Clear thermal hysteresis is observed around $T_{\rm M2}$ between the cooling and warming processes, suggesting a first-order character of the lower transition.

The low-temperature specific heat $C(T)$ for Ce$_5$$M$Ge$_2$ ($M$ = Co, Rh, Ir) is presented in Fig. \ref{figure2}(c). Ce$_5$CoGe$_2$ and Ce$_5$RhGe$_2$ exhibit a characteristic $\lambda$-type anomaly at $T_{\rm C}$, indicative of a second-order magnetic transition. In contrast, Ce$_5$IrGe$_2$ displays two anomalies corresponding to the transitions observed in the resistivity. The higher-temperature anomaly at $T_{\rm M1}$ exhibits a pronounced peak, indicating a second-order magnetic transition, whereas the weaker anomaly at $T_{\rm M2}$ suggests a change of the magnetic state at lower temperature.
Fitting the paramagnetic-regime data with $C(T) = \gamma_0 T + \beta T^{3}$ yields the electronic specific heat coefficients summarized in Table \ref{table3}. 
The resulting $\gamma_0$ values increase progressively from 172.8 mJ$\cdot$K$^{-2}$mol$^{-1}$Ce$^{-1}$ for Ce$_5$CoGe$_2$ to 214.1 mJ$\cdot$K$^{-2}$mol$^{-1}$Ce$^{-1}$ for Ce$_5$RhGe$_2$, and further to 288.7 mJ$\cdot$K$^{-2}$mol$^{-1}$Ce$^{-1}$ for Ce$_5$IrGe$_2$.

Figure \ref{figure2}(d) displays the temperature dependence of the magnetic susceptibility $\chi(T)$ of Ce$_5$$M$Ge$_2$ measured under an external field of 0.05 T, where both zero-field-cooled (ZFC) and field cooled (FC) curves are shown.
The $\chi(T)$ curves for Ce$_5$CoGe$_2$ and Ce$_5$RhGe$_2$ exhibit sharp increases and pronounced ZFC--FC splitting below their ordering temperatures, consistent with FM ground states. 
For Ce$_5$IrGe$_2$, two successive anomalies are observed at $T_{\rm M1}$ and $T_{\rm M2}$. The cusp-like feature at $T_{\rm M1}$ suggests the onset of AFM-like ordering.

\begin{figure*}[htbp]
	\begin{center}
		\includegraphics[width=0.9\textwidth]{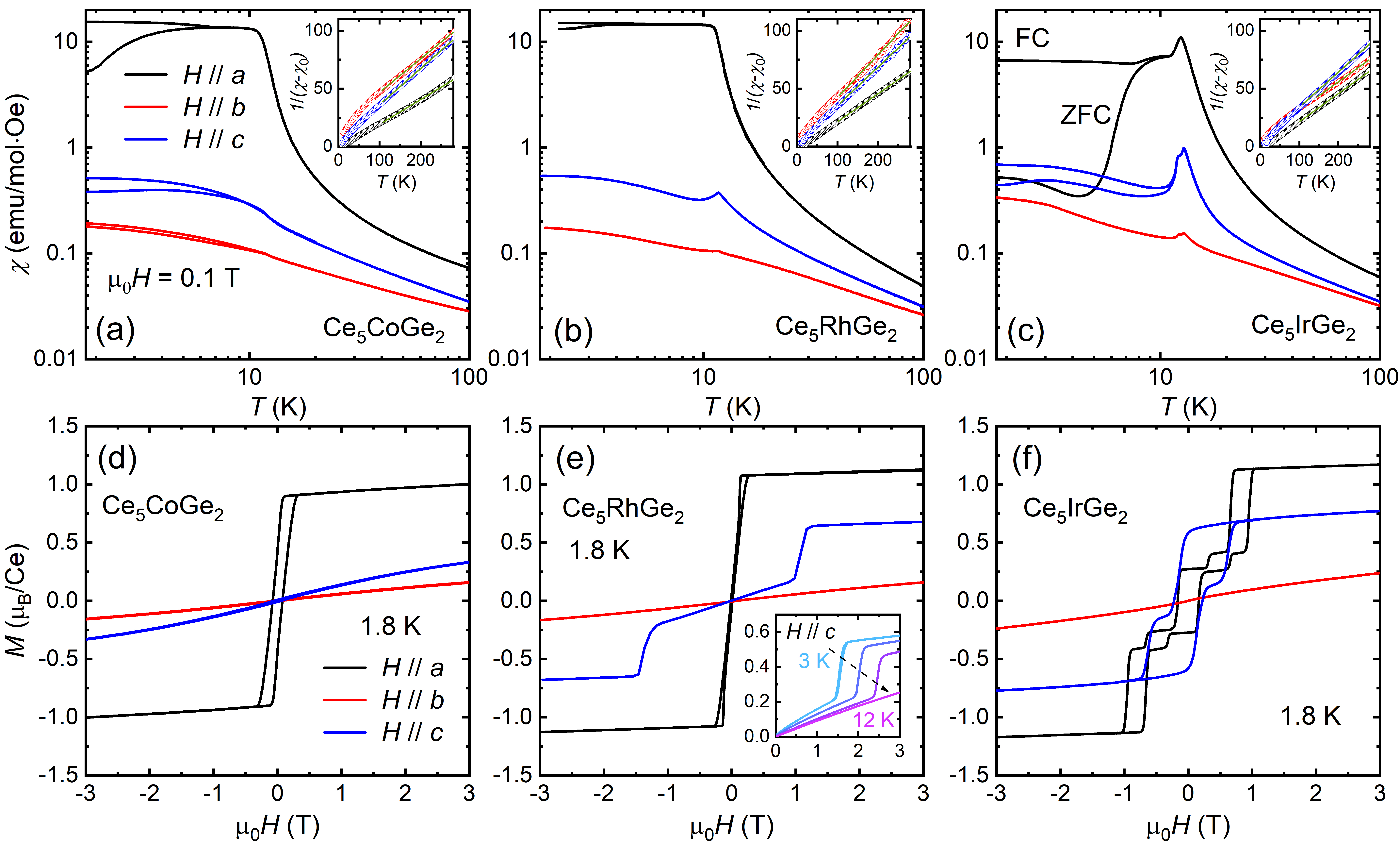}
	\end{center}
	\caption{(Color online) (a)–(c) Temperature dependence of the magnetic susceptibility of Ce$_5$$M$Ge$_2$ ($M$ = Co, Rh, Ir) measured at 0.1 T along different crystallographic orientations. Both ZFC and FC curves are shown. The insets show the $1/(\chi-\chi_0)$ as a function of temperature, where the solid lines correspond to fits to the modified Curie-Weiss law. (d)–(f) The field dependence of the magnetization for the corresponding directions at 1.8 K. The inset of (e) shows $M(H)$ of Ce$_5$RhGe$_2$ measured at 3 K, 5 K, 7 K and 12 K for $H \Vert c$. The dashed arrow denotes the trend of the metamagnetic transition.}
	\label{figure3}
\end{figure*}

The magnetic susceptibility of Ce$_5$$M$Ge$_2$ was measured along the three principal crystallographic directions, as shown in Figs. \ref{figure3}(a)-(c). All three compounds exhibit pronounced uniaxial anisotropy, with $\chi_a(T)$ consistently larger than $\chi_b(T)$ and $\chi_c(T)$ over a broad temperature range, indicating that the $a$ axis is the easy magnetization direction. 
The susceptibility curves show distinct responses near the magnetic transitions across the series. For Ce$_5$CoGe$_2$, all three crystallographic directions exhibit an upward deviation at $T_{\rm C}$, whereas Ce$_5$RhGe$_2$ displays weak peaks in the $b$- and $c$-axis susceptibilities near $T_{\rm C}$. In contrast, sharp peaks are observed along all three directions for Ce$_5$IrGe$_2$.

As shown in the insets of Figs. \ref{figure3}(a)-(c), the high-temperature $\chi(T)$ along each principal axis were fitted using the modified Curie--Weiss law, $\chi(T) = \chi_0 + C/(T - \theta_{\rm p})$. The extracted paramagnetic Curie temperatures $\theta_{\rm p}$ and effective moments $\mu_{\rm eff}$ are summarized in Table \ref{table3}. 
Except for the weakly negative $\theta_p$ value of Ce$_5$CoGe$_2$ ($-2.24$ K) along the $a$ axis, positive $\theta_p$ values are obtained for Ce$_5$RhGe$_2$ ($13.16$ K) and Ce$_5$IrGe$_2$ ($12.20$ K), whereas negative values are observed along the $b$ and $c$ axes for all compounds. 
The extracted effective moments are close to the theoretical value expected for localized Ce$^{3+}$ ions, indicating that the magnetic properties are dominated by Ce $4f$ moments. The pronounced anisotropy of the Curie--Weiss temperatures may reflect the combined influence of CEF effects and anisotropic magnetic interactions in the Ce$_5$$M$Ge$_2$ series.

Figures \ref{figure3}(d)–(f) show the isothermal magnetization curves of Ce$_5$$M$Ge$_2$ ($M$ = Co, Rh, Ir) measured at 1.8 K along the three principal crystallographic directions. Consistent with the susceptibility anisotropy, all three compounds exhibit the largest magnetization for $H \parallel a$, confirming that the $a$ axis is the easy magnetization direction.
For Ce$_5$RhGe$_2$, the $M(H)$ curve along the $a$ axis exhibits a clear hysteresis loop around zero field and rapidly approaches near saturation without any detectable metamagnetic transition, suggesting a FM ground state similar to that of Ce$_5$CoGe$_2$ \cite{dajun2024,pikul2014magnetic}. When the magnetic field is applied along the $c$ axis, a metamagnetic transition is observed, indicating a field-induced realignment of the ordered moments. The critical field of this transition shifts to higher fields with increasing temperature [see the inset of Fig. \ref{figure3}(e)].

Ce$_5$IrGe$_2$ exhibits a distinct magnetic response compared with the Co and Rh materials. For $H \parallel a$, the low-temperature $M(H)$ curve shows a clear hysteresis loop around zero field, indicating that the magnetic ground state at 1.8 K possesses a finite magnetization. However, the low-field magnetization plateaus at a value of approximately $0.25~\mu_{\rm B}$/Ce, which is considerably smaller than the high-field saturation magnetization of  $M_{\rm s} = 1.23~\mu_{\rm B}$/Ce. This reduced value suggests a FIM-like ground state rather than a FM state, with the low-field magnetization corresponding approximately to $M_{\rm s}/5$.
At higher magnetic fields, successive metamagnetic transitions (MM 2 and 3) occur at $\mu_0H_{\rm 1}=0.36$ T and $\mu_0H_{\rm 2}=0.7$ T, resulting in an intermediate magnetization plateau near $M_{\rm s}/3$ before the system reaches the spin-polarized state. 
These fractional magnetization states indicate successive field-induced rearrangements of the magnetic structure in Ce$_5$IrGe$_2$. Hysteretic behavior with a metamagnetic transition is also observed for $H \parallel c$, suggesting, as in Ce$_5$RhGe$_2$, a field-induced reorientation of the ordered moments.

\begin{figure}[htbp]
	\begin{center}
		\includegraphics[width=\columnwidth]{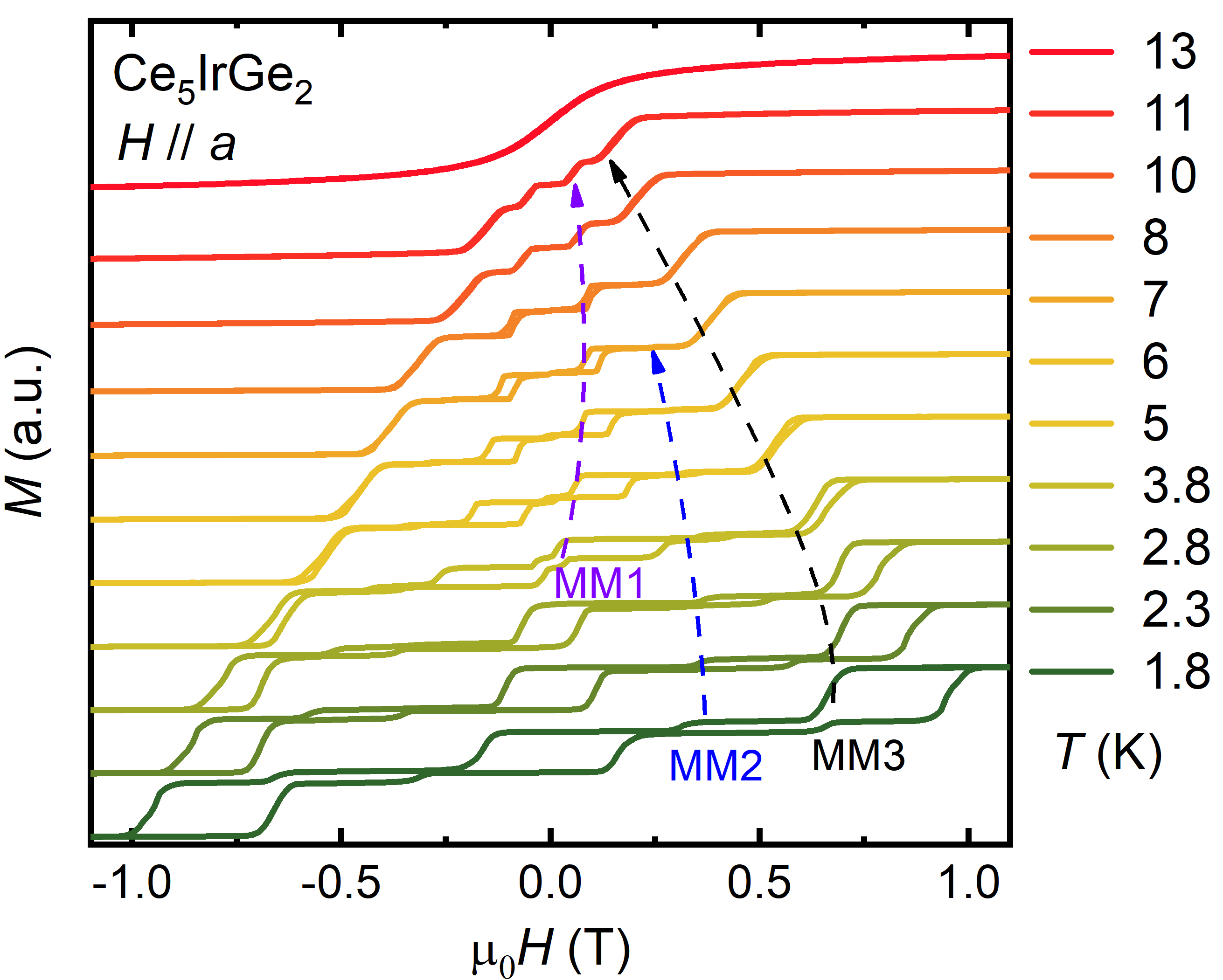}
	\end{center}
	\caption{(Color online) Field dependence of the magnetization of Ce$_5$IrGe$_2$ at various temperatures for $H \parallel a$. The curves are offset vertically for clarity. The dashed arrows denote the trend of the metamagnetic transitions.}
	\label{figure4}
\end{figure}

The multiple metamagnetic transitions and fractional magnetization plateaus observed in Ce$_5$IrGe$_2$ motivate a systematic investigation of their field--temperature evolution for $H \parallel a$. Figure~\ref{figure4} displays the isothermal $M(H)$ curves measured at selected temperatures. 
At the lowest temperatures, Ce$_5$IrGe$_2$ exhibits a pronounced hysteresis loop around zero field, consistent with a FIM-like ground state. With increasing temperature, the zero-field hysteresis progressively narrows. Above approximately 3.8 K, an additional low-field metamagnetic transition, denoted MM 1, emerges and is accompanied by pronounced hysteresis. 
Between approximately 3.8 and 8 K, three successive metamagnetic transitions, MM 1, MM 2, and MM 3, are observed with increasing field. As indicated by the dashed arrows in Fig.~\ref{figure4}, both MM 2 and MM 3 shift continuously toward lower fields with increasing temperature. Meanwhile, MM 2 gradually weakens and disappears above 8 K, where the hysteresis loop is closed about zero-field, and the magnetization corresponds to that of an AFM state. At higher temperatures, only MM 1 and MM 3 remain clearly resolved, with the $M_{\rm s}/3$ magnetization plateau stabilized between them. These results demonstrate that the low-temperature FIM-like ground state does not emerge directly below $T_{\rm M2}$. 
Instead, $T_{\rm M2}$ appears to be a transition into an AFM phase, which at lower temperatures transitions into a FIM-like ground state below around 3.8 K with a $M_{\rm s}/5$ magnetization. Note that a greatly reduced hysteresis loop about zero-field can still be detected between 3.8 K and 8 K. While this could evidence an additional magnetic phase with a small magnetization, it could also signal a coexistence between the AFM phase and low-temperature FIM across this range.

\begin{figure*}[htbp]
	\begin{center}
		\includegraphics[width=0.9\textwidth]{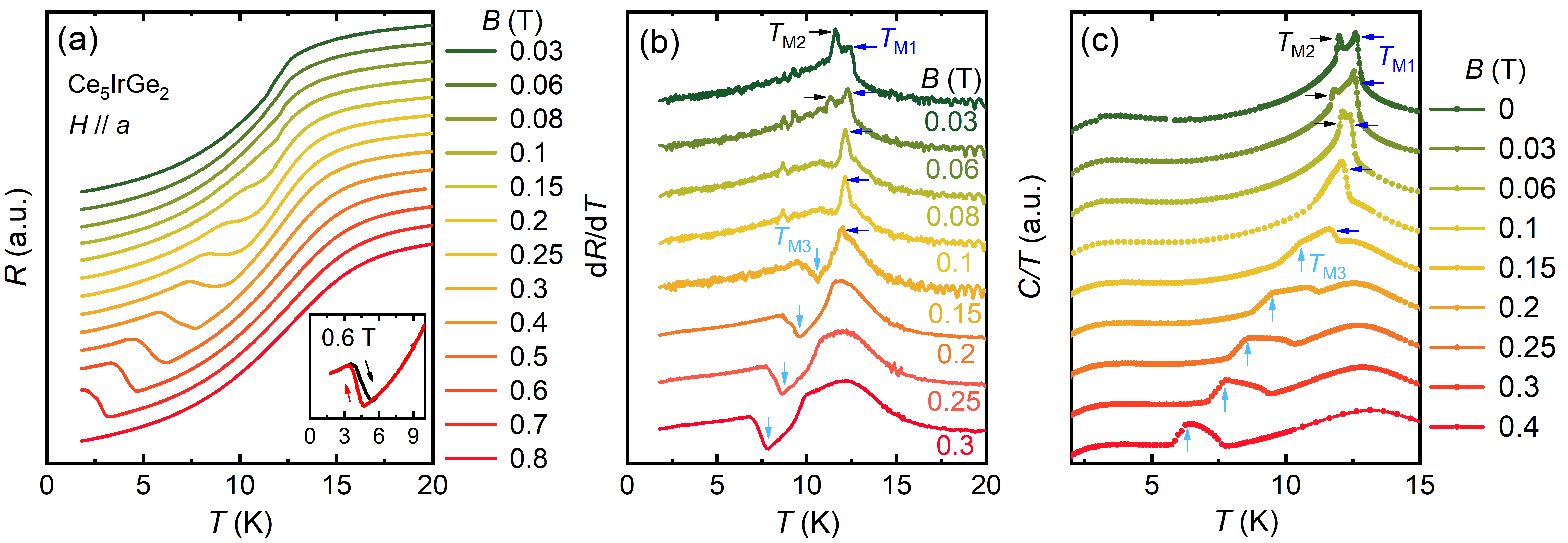}
	\end{center}
	\caption{(Color online) (a) Temperature dependence of the resistance of Ce$_5$IrGe$_2$ in various fields applied along the $a$ axis, with the current applied in the $bc$ plane. The inset shows the thermal hysteresis across $T_{\rm M3}$ at 0.6 T upon heating and cooling. (b) Temperature derivative of resistance $dR/dT$. Black, dark blue, and light blue arrows correspond to $T_{\rm M1}$, $T_{\rm M2}$, and the field-induced transition $T_{\rm M3}$, respectively. (c) Temperature dependence of the specific heat of Ce$_5$IrGe$_2$ in both zero field and different fields applied along the $a$ axis. Curves are offset vertically for clarity.}
	\label{figure5}
\end{figure*}

Figure~\ref{figure5}(a) displays the low-temperature resistance of Ce$_5$IrGe$_2$ measured under various magnetic fields applied along the $a$ axis. With increasing field, the resistance anomalies associated with the magnetic transitions are progressively modified. Above 0.15 T, the resistance anomaly changes from a sharp decrease to an upturn, marking a pronounced change in the character of the magnetic transition. The upturn may indicate a field-induced partial gapping of the Fermi surface \cite{Fisk1991,Zhang_2024}.
As shown in the inset of Fig.~\ref{figure5}(a), a clear thermal hysteresis is observed across the upturn at 0.6 T between the cooling and warming processes.
The temperature derivative $dR/dT$ [Fig.~\ref{figure5}(b)] allows the transition temperatures to be tracked more clearly. At low fields, the anomalies associated with $T_{\rm M1}$ and $T_{\rm M2}$ appear as distinct peaks. Above approximately 0.2 T, the field-induced anomaly denoted as $T_{\rm M3}$ develops and appears as a broad minimum in $dR/dT$. These results are consistent with the low-temperature magnetic state being changed by relatively small applied magnetic fields.

The specific heat, plotted as $C/T$, measured under various magnetic fields for $H\parallel a$ is shown in Fig.~\ref{figure5}(c). For $\mu_0H\gtrsim0.15$ T, the shape of the transition anomaly changes abruptly, in agreement with the evolution observed in the resistivity. Above 0.2 T, a broad maximum develops near the original magnetic ordering temperature and shifts slightly toward higher temperatures with increasing field. This broad maximum may originate from the field-induced splitting of the CEF ground-state doublet. Meanwhile, the low-temperature anomaly associated with $T_{\rm M3}$ becomes progressively more pronounced and shifts toward lower temperatures. The pronounced changes in the profiles of the specific heat anomalies provide thermodynamic evidence for a field-induced change of the magnetic order.

\begin{table*}[!ht]
	\renewcommand\arraystretch{1.4}
	\centering
	\caption{A summary of the physical properties of Ce$_5$$M$Ge$_2$ ($M$ = Co, Rh, Ir): type of magnetism, magnetic ordering temperatures $T_{\rm ord}$ determined from $\chi$(T); Sommerfeld coefficient $\gamma_0$; easy magnetization direction; Curie-Weiss temperatures $\theta_{\rm p}$ and effective moment $\mu_{\rm eff}$ for $H \parallel a$, $H \parallel b$, and $H \parallel c$.}
	\begin{tabular}{ccccccccccc}
		\hline
		~Compound~ & ~Ground state~ & ~$T_{\rm ord}$ (K)~ & ~$\gamma_0$ (mJ$\cdot$K$^{-2}$mol$^{-1}$Ce$^{-1}$)~ & ~Easy-direction~ & ~~$\theta_{\rm p}^{a}$~~ & ~~$\mu_{\rm eff}^{a}$~~ & ~~$\theta_{\rm p}^{b}$~~ & ~~$\mu_{\rm eff}^{b}$~~ & ~~$\theta_{\rm p}^{c}$~~ &  ~~$\mu_{\rm eff}^{c}$~~
		\\ \hline
		Ce$_5$CoGe$_2$ & FM & 11 & 172.8 & $a$ & -2.24 & 2.87 & -67.56 & 2.38 & -14.04 & 2.25 
		\\
		Ce$_5$RhGe$_2$ & FM & 11.5 & 214.1 & $a$ & 13.16 & 2.55 & -27.26 & 2.16 & -5.64 & 2.20 
		\\
		Ce$_5$IrGe$_2$ & FIM & 12.7, 11.8 & 288.7 & $a$ & 12.20 & 2.56 & -48.29 & 2.68 & -9.94 & 2.30 
		\\ \hline
	\end{tabular}
	\label{table3}
\end{table*}

\begin{figure}[htbp]
	\begin{center}
		\includegraphics[width=\columnwidth]{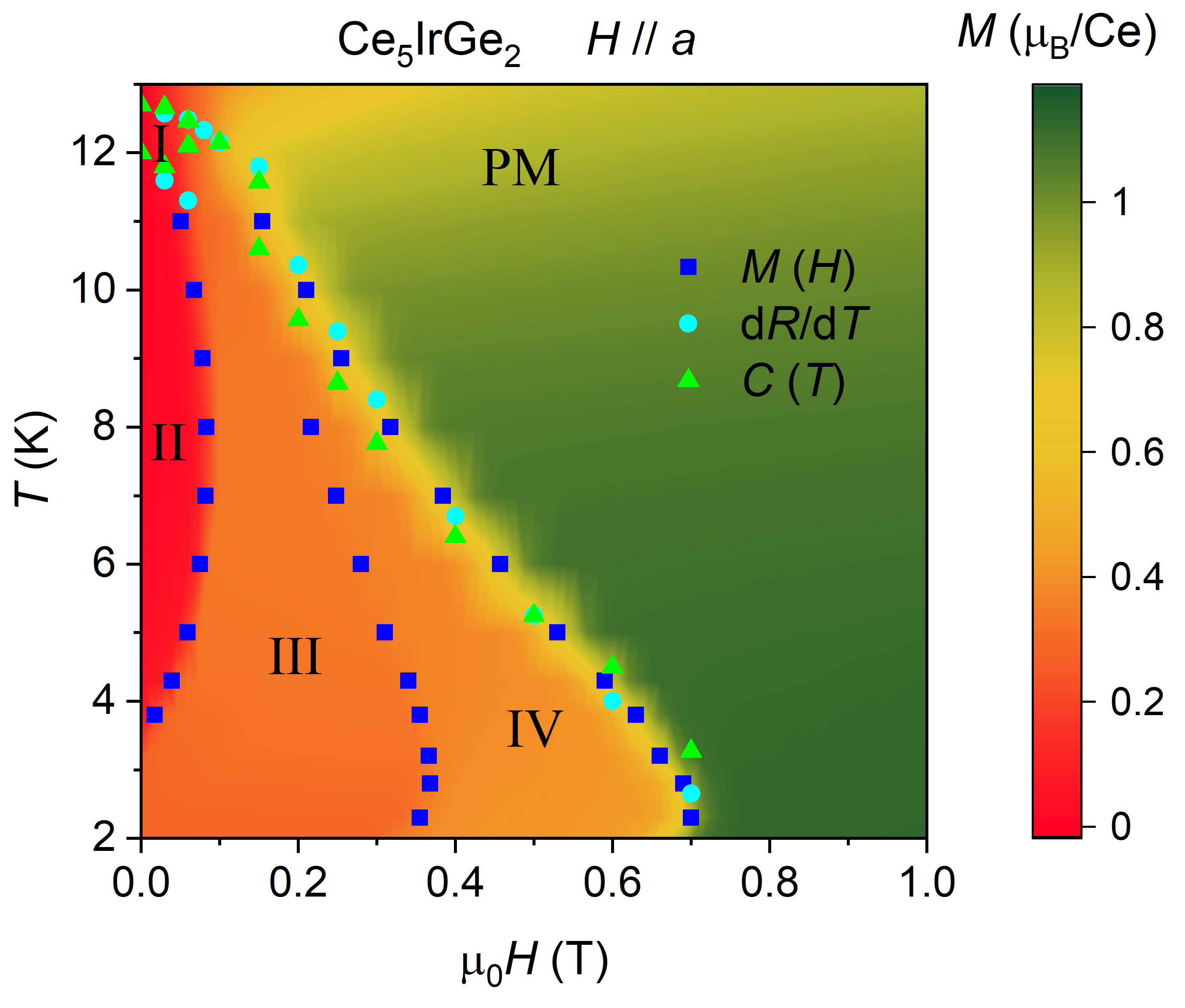}
	\end{center}
	\caption{(Color online) Magnetic field-temperature phase diagram of Ce$_5$IrGe$_2$ for $H \parallel a$. Both the magnetization color map and the phase-boundary points derived from $M(H)$ are obtained from measurements performed upon decreasing the field. The other symbols correspond to transitions determined from $dR/dT$ and $C(T)$, as indicated in the legend.}
	\label{figure6}
\end{figure}

By combining the resistivity, specific heat, and magnetization measurements, we construct the $H$-$T$ phase diagram of Ce$_5$IrGe$_2$ for $H\parallel a$, as shown in Fig.~\ref{figure6}. The phase boundaries determined from different experimental probes show good consistency.
At zero field, Ce$_5$IrGe$_2$ undergoes a magnetic transition at $T_{\rm M1} = 12.7$ K from the paramagnetic state into an AFM-like ordered phase (phase I). A second anomaly at $T_{\rm M2} = 11.8$ K indicates a change of the magnetic state, giving rise to an intermediate ordered phase (phase II).
Below approximately 4 K, the system enters a FIM ground state with a $M_{\rm s}/5$ magnetization and a pronounced hysteresis loop about zero-field (phase III). Under applied magnetic fields, phase III extends to higher temperatures and is separated from phase II by the low-field metamagnetic transition MM 1 between approximately 3.8 and 8 K. At higher fields, MM 2 separates phase III from phase IV, where the latter is associated with the $M_{\rm s}/3$ magnetization plateau, while MM 3 marks the boundary between phase IV and the high-field spin-polarized state.
A notable feature of the phase diagram is the strong temperature dependence of the low-field magnetic phase. Upon cooling, phase III progressively expands toward zero magnetic field, consistent with the development of the zero-field hysteresis observed in the magnetization measurements.

\section{Discussion and Summary}

Despite sharing the same crystal structure, the Ce$_5M$Ge$_2$ ($M$ = Co, Rh, Ir) series exhibit a range of distinct magnetic ground states, with Ce$_5$CoGe$_2$ and Ce$_5$RhGe$_2$ showing closely related FM behavior, whereas Ce$_5$IrGe$_2$ exhibits multiple magnetic phases.

Neutron diffraction studies have established that Ce$_5$CoGe$_2$ possesses a non-collinear FM ground state rather than a simple collinear FM structure \cite{Wu2026}. The magnetic response of Ce$_5$RhGe$_2$ closely resembles that of Ce$_5$CoGe$_2$, as both compounds exhibit hysteresis loops along the easy magnetization axis without additional metamagnetic transitions. These similarities suggest that Rh substitution likely preserves a comparable non-collinear FM ground state. However, the additional metamagnetic transition observed for $H\parallel c$ in Ce$_5$RhGe$_2$ indicates weaker magnetocrystalline anisotropy compared with Ce$_5$CoGe$_2$, allowing a field-induced rearrangement of the magnetic moments along the hard axis while maintaining a similar zero-field magnetic configuration.

In contrast, Ce$_5$IrGe$_2$ exhibits a qualitatively different magnetic evolution from the Co and Rh analogues, suggesting that transition-metal substitution significantly alters the balance of competing magnetic interactions in this family. FIM ground states accompanied by field-induced metamagnetic transitions have also been reported in Ce$_5$PdGe$_2$ and Ce$_5$RuGe$_2$ \cite{skornia2017electronic,skornia2018electronic}, indicating that ferrimagnetism and metamagnetic behavior occur more broadly within the Ce$_5M$Ge$_2$ family. Similar fractional plateaus, including the $M_{\rm s}/3$ plateau observed here, have been reported in other complex magnetic systems such as the frustrated Ising-chain compound Ca$_3$Co$_2$O$_6$ \cite{AaslandS1997} and the rare-earth intermetallic $R$Rh$_6$Ge$_4$ ($R$ = Tb - Ho) \cite{ChenY2023,Zhang2025}, where competing magnetic interactions can stabilize distinct field-induced magnetic configurations.
In Ce$_5$CoGe$_2$, the four crystallographically inequivalent Ce sites carry ordered moments of different magnitudes, while their net moments remain coaligned \cite{Wu2026}. The coexistence of AFM-like phases, fractional magnetization plateaus, and an uncompensated low-temperature moment in Ce$_5$IrGe$_2$ suggests a substantially different arrangement of the Ce moments. Although its microscopic magnetic structure remains to be determined, one possible scenario is that the moments on one or more inequivalent Ce sites align antiparallel to those on the remaining Ce sublattices. Such a partially compensated configuration could account for the FIM-like ground state and provide a natural origin for the multiple field-induced magnetic phases.

In summary, we have systematically investigated the structural, transport, thermodynamic, and magnetic properties of single-crystalline Ce$_5M$Ge$_2$ ($M$ = Co, Rh, Ir). All three compounds crystallize in the orthorhombic \emph{Pnma} structure and exhibit a common easy magnetization axis along the crystallographic $a$ direction. Ce$_5$CoGe$_2$ and Ce$_5$RhGe$_2$ undergo FM ordering with Curie temperatures of approximately 11 K and 11.5 K, respectively. In contrast, Ce$_5$IrGe$_2$ exhibits two successive magnetic transitions at $T_{\rm M1} = 12.7$ K and $T_{\rm M2} = 11.8$ K, together with multiple metamagnetic transitions giving rise to fractional magnetization plateaus at approximately $M_{\rm s}/5$ and $M_{\rm s}/3$. 
The constructed $H$-$T$ phase diagram further reveals that the critical field of the low-field metamagnetic transition is suppressed upon cooling. At the lowest temperatures, the appearance of a pronounced zero-field hysteresis together with the $M_{\rm s}/5$ magnetization establishes a FIM-like ground state in Ce$_5$IrGe$_2$.

As a result, the ambient-pressure characterization of the Ce$_5M$Ge$_2$ ($M$ = Co, Rh, Ir) series provides an essential foundation for understanding the evolution of magnetic ground states and other pressure-induced phenomena. Further neutron diffraction experiments on Ce$_5$IrGe$_2$ will be crucial for determining its microscopic magnetic structures and identifying the magnetic configurations responsible for the fractional magnetization states. In addition, extending high-pressure studies to Ce$_5$RhGe$_2$ and Ce$_5$IrGe$_2$ will clarify whether the quantum critical behavior and superconductivity observed in Ce$_5$CoGe$_2$ \cite{Zhangyn2026} represent a universal feature of this family or depend sensitively on the underlying magnetic ground state.

\section{acknowledgments}
This work was supported by the National Key R$\&$D Program of China (No. 2022YFA1402200 and No. 2023YFA1406303), the National Natural Science Foundation of China (No. W2511006, No. U23A20580, No. 12674175, No. 12550401 and No. 12494592), the Zhejiang Provincial Natural Science Foundation (No. LRG26A040001), and the New Cornerstone Science Foundation (No. NCI202509).

\bibliography{ref}

@article{AaslandS1997,
  title = {{Magnetic Properties of the One-Dimensional Ca{$_3$}Co{$_2$}O{$_6$}}},
  author = {Aasland, S. and Fjellv{\aa}g, H. and Hauback, B.},
  year = {1997},
  month = jan,
  journal = {Solid State Commun.},
  volume = {101},
  pages = {187--192},
  issn = {0038-1098},
  doi = {10.1016/S0038-1098(96)00531-5}
}

@article{ChenY2023,
  title = {{Multiple Magnetic Phases and Magnetization Plateaus in TbRh{\textsubscript{6}}Ge{\textsubscript{4}}}},
  author = {Chen, Y. X. and Zhang, Y. J. and Li, Rui and Su, Hang and Shan, Z. Y. and Smidman, M. and Yuan, H. Q.},
  year = {2023},
  month = mar,
  journal = {Phys. Rev. B},
  volume = {107},
  pages = {094414},
  issn = {2469-9950, 2469-9969},
  doi = {10.1103/PhysRevB.107.094414}
}

@article{Si2010,
  title = {{Heavy Fermions and Quantum Phase Transitions}},
  author = {Si, Qimiao and {F. Steglich}},
  year = {2010},
  journal = {Science (New York, N.Y.)},
  volume = {329},
  pages = {1161--1166},
  doi = {10.1126/science.1191195}
}

@article{GegenwartP2008,
  title = {{Quantum Criticality in Heavy-Fermion Metals}},
  author = {Gegenwart, Philipp and Si, Qimiao and Steglich, Frank},
  year = {2008},
  month = mar,
  journal = {Nat. Phys.},
  volume = {4},
  pages = {186--197},
  issn = {1745-2481},
  doi = {10.1038/nphys892}
}

@article{pikul2014magnetic,
  title = {{Magnetic and Related Properties of Ce{\textsubscript{5}}CoGe{\textsubscript{2}}, CeCoGe and CeCo{\textsubscript{2}}Ge{\textsubscript{2}}}},
  author = {Pikul, A. and Pasturel, Mathieu and Wi{\'s}niewski, Piotr and Soud{\'e}, A and Tougait, Olivier and No{\"e}l, Henri and Kaczorowski, Dariusz},
  year = {2014},
  journal = {Intermetallics},
  volume = {53},
  pages = {40--44},
  publisher = {Elsevier},
  doi = {10.1016/j.intermet.2014.03.021}
}

@article{Brando2016,
  title = {{Metallic Quantum Ferromagnets}},
  author = {Brando, M. and Belitz, D. and Grosche, F. M. and Kirkpatrick, T. R.},
  year = {2016},
  month = may,
  journal = {Rev. Mod. Phys.},
  volume = {88},
  pages = {025006},
  publisher = {American Physical Society},
  doi = {10.1103/RevModPhys.88.025006}
}

@article{ShenB2020,
  title = {{Strange-Metal Behaviour in a Pure Ferromagnetic Kondo Lattice}},
  author = {Shen, Bin and Zhang, Y. J. and Komijani, Yashar and Nicklas, Michael and Borth, Robert and Wang, An and Chen, Ye and Nie, Z. Y. and Li, Rui and Lu, Xin and Lee, Hanoh and Smidman, Michael and Steglich, Frank and Coleman, Piers and Yuan, H. Q.},
  year = {2020},
  month = mar,
  journal = {Nature},
  volume = {579},
  pages = {51--55},
  issn = {1476-4687},
  doi = {10.1038/s41586-020-2052-z}
}

@article{skornia2017electronic,
  title = {{Electronic Structure, Magnetic, Electric Transport, and Thermal Properties of Ce{\textsubscript{5}}PdGe{\textsubscript{2}}}},
  author = {Skornia, Pawe{\l} and Goraus, Jerzy and Fija{\l}kowski, Marcin and {\'S}lebarski, Andrzej},
  year = {2017},
  journal = {J. Alloys and Compd.},
  volume = {724},
  pages = {222--228},
  publisher = {Elsevier},
  doi = {10.1016/j.jallcom.2017.06.304}
}

@article{skornia2018electronic,
  title = {{Electronic Structure and Magnetic Properties of the Magnetically Ordered Intermediate Valent Ce{\textsubscript{5}}RuGe{\textsubscript{2}}}},
  author = {Skornia, Pawe{\l} and Goraus, Jerzy and Fija{\l}kowski, Marcin and {\'S}lebarski, Andrzej},
  year = {2018},
  journal = {J. Alloys and Compd.},
  volume = {767},
  pages = {512--521},
  publisher = {Elsevier},
  doi = {10.1016/j.jallcom.2018.07.076}
}

@article{skornia2019magnetic,
  title = {{Magnetic and Electrical Transport Properties of Ce{\textsubscript{5}}RhGe{\textsubscript{2}}}},
  author = {Skornia, Pawe{\l} and Deniszczyk, Jozef and Fija{\l}kowski, Marcin and {\'S}lebarski, Andrzej},
  year = {2019},
  journal = {J. Alloys and Compd.},
  volume = {775},
  pages = {524--532},
  publisher = {Elsevier},
  doi = {10.1016/j.jallcom.2018.10.096}
}

@article{sologub2000formation,
  title = {{Formation, Crystal Structure and Magnetism of Ternary Compounds Ce{\textsubscript{5}}{\emph{M}}Ge{\textsubscript{2}} ({\emph{M}}= Co, Ni, Ru, Rh, Pd, Ir, Pt)}},
  author = {Sologub, O. L. and Salamakha, P. S. and Godart, C.},
  year = {2000},
  journal = {J. Alloys and Compd.},
  volume = {307},
  pages = {31--39},
  publisher = {Elsevier},
  doi = {10.1016/S0925-8388(00)00824-0}
}

@article{weng2016multiple,
  title = {{Multiple Quantum Phase Transitions and Superconductivity in Ce-Based Heavy Fermions}},
  author = {Weng, Z. F. and Smidman, M and Jiao, L and Lu, Xin and Yuan, H. Q.},
  year = {2016},
  journal = {Rep. Prog. Phys.},
  volume = {79},
  pages = {094503},
  publisher = {IOP Publishing},
  doi = {10.1088/0034-4885/79/9/094503}
}

@article{Zhang2025,
  title = {{Magnetic Properties of {\emph{R}}Rh{\textsubscript{6}}Ge{\textsubscript{4}} ({\emph{R}}=Pr, Nd, Sm, Gd--Er) Single Crystals}},
  author = {Zhang, J. W. and Zhang, Y. J. and Chen, Y. X. and Shan, Z. Y. and Zhan, Jin and Wang, M. Y. and Liu, Yu and Smidman, Michael and Yuan, H. Q.},
  year = {2025},
  month = oct,
  journal = {Phys. Rev. B},
  volume = {112},
  pages = {134454},
  publisher = {American Physical Society},
  doi = {10.1103/x8nk-gq44}
}

@article{Wu2026,
	title = {{Noncollinear Ferromagnetism in the Kondo-Lattice Compound Ce{\textsubscript{5}}CoGe{\textsubscript{2}}}},
	author = {Wu, J. Y. and Zhang, J. W. and Shiroka, Toni and Islam, Shams Sohel and Wang, M. Y. and Zhang, Y. J. and Adroja, Devashibhai T. and Liu, Yu and Yuan, H. Q. and Smidman, Michael},
	year = {2026},
	month = apr,
	journal = {Phys. Rev. B},
	volume = {113},
	pages = {134431},
	publisher = {American Physical Society},
	issn = {2469-9950, 2469-9969},
	doi = {10.1103/r1qt-9pkk}
}

@article{Zhangyn2026,
	title = {{Pressure-Induced Superconductivity beyond Magnetic Quantum Criticality in a Kondo Ferromagnet}},
	author = {Zhang, Y. N. and Zhang, Y. J. and Zhang, J. W. and Ye, K. X. and Su, D. J. and Huang, Y. E. and Shan, Z. Y. and Li, J. Y. and Li, Rui and Chen, Ye and Lu, Xin and Jiao, Lin and Liu, Yu and Smidman, Michael and Steglich, Frank and Yuan, H. Q.},
	year = {2026},
	month = feb,
	journal = {Natl. Sci. Rev.},
	volume = {13},
	pages = {nwag119},
	issn = {2095-5138, 2053-714X},
	doi = {10.1093/nsr/nwag119}
}

@article{dajun2024,
	title = {{Coexistence of Ferromagnetism and Cluster Glass Behavior in Ce{\textsubscript{5}}CoGe{\textsubscript{2}}}},
	author = {Su, D. J. and Zhang, J. W. and Zhang, Y. J. and Shan, Z. Y. and Zhang, Y. N. and Smidman, Michael and Jiao, Lin and Liu, Yu and Yuan, H. Q.},
	year = {2024},
	month = oct,
	journal = {Phys. Rev. B},
	volume = {110},
	pages = {144432},
	publisher = {American Physical Society},
	issn = {2469-9950, 2469-9969},
	doi = {10.1103/PhysRevB.110.144432}
}

@article{Zhan2025,
	title = {{Critical Fluctuations and Conserved Dynamics in a Strange Ferromagnetic Metal}},
	author = {Zhan, Jin and Zhang, Y. J. and Zhang, J. W. and Liu, Yu and Nie, Z. Y. and Chen, Y. X. and Jiao, Lin and Komijani, Yashar and Smidman, Michael and Steglich, Frank and Coleman, Piers and Yuan, H. Q.},
	year = {2025},
	month = dec,
	journal = {Phys. Rev. Lett.},
	volume = {135},
	pages = {266504},
	publisher = {American Physical Society},
	doi = {10.1103/bp6l-46z7}
}

@article{Fisk1991,
	title = {{Massive Electron State in YbBiPt}},
	author = {Fisk, Z. and Canfield, P. C. and Beyermann, W. P. and Thompson, J. D. and Hundley, M. F. and Ott, H. R. and Felder, E. and Maple, M. B. and Lopez De La Torre, M. A. and Visani, P. and Seaman, C. L.},
	year = {1991},
	month = dec,
	journal = {Phys. Rev. Lett.},
	volume = {67},
	pages = {3310--3313},
	publisher = {American Physical Society},
	issn = {0031-9007},
	doi = {10.1103/PhysRevLett.67.3310}
}

@article{Zhang_2024,
	title = {{Structural and Physical Properties of the Heavy Fermion Metal Ce{\textsubscript{2}}NiAl{\textsubscript{6}}Si{\textsubscript{5}}}},
	author = {Zhang, J. W. and Wu, J. W. and Chen, Ye and Li, Rui and Smidman, Michael and Liu, Yu and Song, Yu and Yuan, H. Q.},
	year = {2024},
	month = dec,
	journal = {Chinese Phys. Lett.},
	volume = {41},
	pages = {127304},
	publisher = {{Chinese Physical Society and IOP Publishing Ltd}},
	issn = {0256-307X, 1741-3540},
	doi = {10.1088/0256-307X/41/12/127304}
}

@article{Uhlarz2004,
	title = {{Quantum Phase Transitions in the Itinerant Ferromagnet ZrZn{\textsubscript{2}}}},
	author = {Uhlarz, M and Pfleiderer, C and Hayden, S. M.},
	year = {2004},
	journal = {Phys. Rev. Lett.},
	volume = {93},
	pages = {256404},
	publisher = {APS},
	doi = {10.1103/PhysRevLett.93.256404}
}

@article{FriedemannS2018,
	title = {{Quantum Tricritical Points in NbFe$_2$}},
	author = {Friedemann, Sven and Duncan, Will J. and Hirschberger, Max and Bauer, Thomas W. and K{\"u}chler, Robert and Neubauer, Andreas and Brando, Manuel and Pfleiderer, Christian and Grosche, F.~Malte},
	year = {2018},
	month = jan,
	journal = {Nat. Phys.},
	volume = {14},
	pages = {62--67},
	issn = {1745-2473, 1745-2481},
	doi = {10.1038/nphys4242}
}

@article{Kote2013,
	title = {{Pressure--Temperature--Magnetic Field Phase Diagram of Ferromagnetic Kondo Lattice CeRuPO}},
	author = {Kotegawa, Hisashi and Toyama, Toshihiro and Kitagawa, Shunsaku and Tou, Hideki and Yamauchi, Ryota and Matsuoka, Eiichi and Sugawara, Hitoshi},
	year = {2013},
	month = dec,
	journal = {J. Phys. Soc. Jpn.},
	volume = {82},
	pages = {123711},
	issn = {0031-9015, 1347-4073},
	doi = {10.7566/JPSJ.82.123711}
}

@article{SteppkeA2013,
	title = {{Ferromagnetic Quantum Critical Point in the Heavy-Fermion Metal YbNi{\textsubscript{4}}(P{\textsubscript{1-x}}As{\textsubscript{x}}){\textsubscript{2}}}},
	author = {Steppke, Alexander and K{\"u}chler, Robert and Lausberg, Stefan and Lengyel, Edit and Steinke, Lucia and Borth, Robert and L{\"u}hmann, Thomas and Krellner, Cornelius and Nicklas, Michael and Geibel, Christoph and Steglich, Frank and Brando, Manuel},
	year = {2013},
	month = feb,
	journal = {Science},
	volume = {339},
	pages = {933--936},
	publisher = {American Association for the Advancement of Science},
	doi = {10.1126/science.1230583}
}

@article{Aoki2014superconductivity,
	title = {{Superconductivity and Ferromagnetic Quantum Criticality in Uranium Compounds}},
	author = {Aoki, Dai and Flouquet, Jacques},
	year = {2014},
	month = jun,
	journal = {J. Phys. Soc. Jpn.},
	volume = {83},
	pages = {061011},
	issn = {0031-9015, 1347-4073},
	doi = {10.7566/JPSJ.83.061011}
}

@article{Wester2009,
	title = {{Kondo-Cluster-Glass State near a Ferromagnetic Quantum Phase Transition}},
	author = {Westerkamp, T. and Deppe, M. and K{\"u}chler, R. and Brando, M. and Geibel, C. and Gegenwart, P. and Pikul, A. P. and Steglich, F.},
	year = {2009},
	month = may,
	journal = {Phys. Rev. Lett.},
	volume = {102},
	pages = {206404},
	publisher = {American Physical Society},
	issn = {0031-9007, 1079-7114},
	doi = {10.1103/PhysRevLett.102.206404}
}

\end{document}